\documentclass[runningheads]{llncs}
\usepackage{graphicx}
\usepackage{url} 
\usepackage{tikz} 
\usepackage{multirow}
\usepackage[misc,geometry]{ifsym}
\usepackage[english]{babel}
\usepackage[autostyle]{csquotes}

\newcommand*\ce[1][1ex]{\tikz\draw (0,0) circle (0.8ex);} %
\newcommand*\ch[1][1ex]{%
  \begin{tikzpicture}
  \draw[fill] (0,0)-- (90:0.8ex) arc (90:270:0.8ex) -- cycle ;
  \draw (0,0) circle (0.8ex);
  \end{tikzpicture}} %
\newcommand*\cf[1][1ex]{\tikz\fill (0,0) circle (0.8ex);} %
\def\tick{\tikz\fill[scale=0.3](0,.35) -- (.25,0) -- (1,.7) -- (.25,.15) -- cycle;} %

\begin{document}
\title{ZLeaks: Passive Inference Attacks on Zigbee based Smart Homes}
\author{Narmeen Shafqat\inst{1}\textsuperscript{\Letter} \and
Daniel J. Dubois\inst{1} \and
David Choffnes\inst{1} \and
Aaron Schulman\inst{2} \and
Dinesh Bharadia\inst{2} \and
Aanjhan Ranganathan\inst{1}}
\authorrunning{N. Shafqat et al.}
\institute{Northeastern University, Boston, MA, USA \\
\email{\{shafqat.n, d.dubois, d.choffnes, aanjhan\}@northeastern.edu} \and
University of California San Diego (UCSD), California, USA \\
\email{schulman@cs.ucsd.edu, dinesh@ucsd.edu}}
\maketitle             
\let\thefootnote\relax\footnotetext{\\This article is accepted at the 20th International Conference on Applied Cryptography and Network Security, ACNS 2022. The conference proceedings are copyrighted by Springer and published in Lecture Notes in Computer Science series. The final authenticated version is available online at https://doi.org/[to-be-assigned].}
\begin{abstract}{
Zigbee is an energy-efficient wireless IoT protocol that is increasingly being deployed in smart home settings. 
In this work, we analyze the privacy guarantees of Zigbee protocol. 
Specifically, we present ZLeaks, a tool that passively identifies in-home devices or events from the encrypted Zigbee traffic by 1) inferring a single application layer (APL) command in the event's traffic, and 2) exploiting the device's periodic reporting pattern and interval. 
This enables an attacker to infer user's habits or determine if the smart home is vulnerable to unauthorized entry. 
We evaluated ZLeaks' efficacy on 19 unique Zigbee devices across several categories and 5 popular smart hubs in three different scenarios; controlled RF shield, living smart-home IoT lab, and third-party Zigbee captures. 
We were able to i) identify unknown events and devices (without a-priori device signatures) using command inference approach with 83.6\% accuracy, ii) automatically extract device's reporting signatures, iii) determine known devices using the reporting signatures with 99.8\% accuracy, and iv) identify APL commands in a public capture with 91.2\% accuracy. 
In short, we highlight the trade-off between designing a low-power, low-cost wireless network and achieving privacy guarantees. 
We have also released ZLeaks tool for the benefit of the research community.
\keywords{Zigbee \and IoT \and Device identification \and Passive inference}
}\end{abstract}
\section{Introduction}
Smart home products (e.g., bulbs, outlets, sensors, etc.) allow users to control and monitor their smart home's environment wirelessly, but unfortunately, pose a significant risk to users' privacy.
Prior studies have demonstrated that by intercepting the IP traffic of a smart home, the attacker can determine in-home devices~\cite{audi,profiliot,devtype1}, events~\cite{event1,pingpong}, and user's habits~\cite{anybodyhome}. 
In practice, these attacks are difficult to carry out, as the attacker must find a vulnerability to capture the user's IP network traffic (e.g., by gaining root access to the home router).
Yet, there exists an easy privacy violation attack, i.e., simply sniffing the Internet of Things (IoT) wireless protocol (e.g., Zigbee) transmissions that are unintentionally emitted to up to hundreds of feet.
Although the IoT traffic is encrypted to prevent eavesdropping, researchers recently showed that the attacker can still identify events using a-priori device signatures~\cite{peekaboo,homonit} and infer a few encrypted Zigbee (Network layer) commands by exploiting the payload lengths~\cite{zigator}. 

In this work, we analyze the privacy guarantees of one of the most popular IoT wireless protocols, Zigbee~\cite{standard}, that is increasingly being used in smart hubs such as Amazon Echo Plus, Samsung SmartThings, and Philips Hue. 
With the launch of more than 500 new Zigbee-certified devices in 2020 alone and the expected sale of nearly four billion Zigbee chipsets by 2023~\cite{z560}, Zigbee continues to be the preferred choice of device manufacturers.

Our key insight is that design optimizations incorporated into Zigbee to enable low-latency communication on low-cost resource-constrained devices fundamentally leak information, e.g., to keep the frame length small, Zigbee performs encryption transformation~\cite{standard} on AES encrypted output to match the message length. %
This enables an eavesdropper to exploit unpadded payload lengths and discrepancies in traffic metadata to infer \textit{every} encrypted network layer (NWK) and application layer (APL) command.
Moreover, to prevent device timeout, Zigbee devices periodically report attributes like battery level, temperature, etc., to the smart hub.   
The distinct reporting patterns and intervals inadvertently serve as device fingerprints. 
In this work, we exploit device's unique reporting patterns and the possibility of inferring APL commands to passively determine devices and events in the target network.
Specifically, we make following contributions.

\textbf{Device and Event Identification using Inferred APL Command: }
We demonstrate that the event traffic of a device always includes at least one functionality-specific APL command (such as \textit{Door Lock/Unlock}), which alone specifies the triggered event (i.e., lock/unlock) and the functional device type (i.e., door lock). 
Zigbee Cluster Library (ZCL) specification~\cite{zcl} inherently leaks information about all such APL commands.
We attempt to infer a single functionality specific APL command in the encrypted event traffic to determine event and device type and combine manufacturer's identity obtained from the Organizationally Unique Identifier (OUI) of the device's MAC address to identify a particular Zigbee device.
Unlike prior works~\cite{peekaboo,homonit}, this approach does not require device's event signatures and can even identify unknown events and devices\footnote{Zigbee Devices not previously observed, i.e., no a-priori access to their traffic.}.

In practice, inferring functionality-specific APL commands is extremely challenging, and so far, no study has attempted it. 
This is because the metadata of functionality-specific APL commands is immensely similar to a hundred other generic APL commands.  
Few APL commands are also manufacturer configurable, which prevent us from exploiting only the payload length, packet direction, and radius (hops) to infer APL commands using prior NWK command inference approach~\cite{zigator}. 
We utilize frame format guidelines~\cite{zcl} to identify all possible APL commands with payload lengths overlapping with the functionality-specific APL commands and their response commands (if any), e.g., \textit{door unlock request and response}. 
The discrepancies in the traffic's metadata, together with the device's logical type (electricity-powered or battery-powered), are used to construct inference rules for each target functionality-specific APL command. 

\textbf{Device Identification using Periodic Reporting Patterns: } 
Zigbee devices periodically report attributes to the smart hub. 
We exploit reporting patterns and intervals to create unique device fingerprints. 
This approach is useful for identifying a known device with unpatched vulnerability (e.g., to spread malware) in the Zigbee network, which has minimal user activity.  
Unlike prior works~\cite{peekaboo,homonit} that analyze Zigbee traffic generated due to event occurrence only: this approach can identify devices even when no event is triggered.
Given that every device's current consumption varies based on its communication pattern and hardware, the periodic reporting time is not trivial to modify as it directly impacts device certification requirement of minimum 2-years battery life~\cite{batteryreq}.

\textbf{Automating event and device identification with ZLeaks tool: } 
We developed a comprehensive privacy analysis tool for Zigbee protocol, named ZLeaks~\cite{zleaks}, that automates the aforementioned identification techniques. ZLeaks takes the Zigbee traffic as input and passively determines events and devices in the smart home. It can also extract devices' reporting signatures automatically.  

We experimentally evaluated ZLeaks on by far the most extensive device set used in privacy analysis of Zigbee protocol including 5 popular smart hubs (SmartThings, Amazon Echo Plus, Philips Hue, OSRAM Lightify, and Sengled) and 27 commercial off-the-shelf Zigbee devices, out of which 19 devices were unique.
The experiments were performed in 1) an isolated RF shield and 2) a living smart-home \enquote{Mon(IoT)r Lab}~\cite{moniotr} with multiple IoT and non-IoT networks operating simultaneously. 
Furthermore, we validated the findings on third-party capture files available on Wireshark~\cite{wshark1} and Crawdad~\cite{crawdad} forums.
Our results indicate that ZLeaks identified event and device information using inferred APL commands with 83.6\% accuracy and devices using reporting patterns with 99.8\% accuracy.
Also, we inferred functionality-specific APL commands in a public Zigbee capture, using our command inference rules, with 91.2\% accuracy.
\section{Background and Motivation}
\subsection{Zigbee Overview}
Zigbee is one of the most popular low-cost, low-power, wireless protocols specifically designed for battery-powered applications in smart ecosystems such as smart homes and industries. 
Zigbee is built on top of the low data-rate IEEE 802.15.4 wireless personal area networking (PAN) standard and implements the physical (PHY) and medium access control (MAC) layers as defined by the IEEE standard.
Most commercial Zigbee devices operate at a data rate of 250 kbps in the 2.4 GHz band (divided into 16 channels, each 5 MHz apart). 
Some Zigbee devices also operate in the unlicensed frequency bands of 784, 868, and 915 MHz.
  
\subsubsection{\textbf{Network Architecture:}}
Zigbee supports both centralized and distributed network architectures. 
Centralized networks comprise of three logical device types; Zigbee coordinator (ZC), Zigbee router (ZR), and Zigbee end-device (ZED), while the distributed networks have ZR and ZED only. 
ZEDs do not route traffic and may sleep to conserve battery, making them appropriate for battery-powered devices (e.g., sensors, door locks). 
ZRs are responsible for routing traffic between nodes and storing messages intended for ZEDs until they are requested.
Every Zigbee network has one ZC that is responsible for network formation, issuing network identifiers, and logical network addresses. 
ZC also acts as a trust center to authenticate new nodes and distribute keys.
ZRs and ZCs are powered devices (e.g., bulbs, smart hubs) and do not sleep during the network's lifetime. 
Besides, Zigbee supports connectivity in star, mesh, and tree topologies. 
Zigbee does not implement MAC address randomization.
Each Zigbee node has a manufacturer-assigned 64-bit MAC (extended) address that is mapped to a unique 16-bit network (logical) address by the ZC during device pairing. 
The logical address is used for routing, while the extended address is used for authentication. 

\begin{figure}[t]
\centering
\includegraphics[scale=0.78]{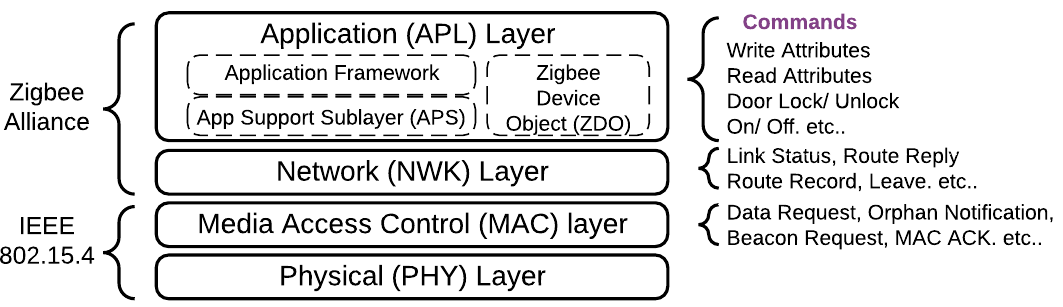}
\caption{Zigbee's Protocol Stack comprising of PHY, MAC, NWK and APL layers }
\label{fig:zigbee_stack}
\end{figure}

\subsubsection{\textbf{Zigbee Protocol Stack (Figure~\ref{fig:zigbee_stack}):}}
Zigbee standard~\cite{standard} defines the functionalities of the Network and Application layers. 
The Network layer is responsible for network formation and management, routing and address allocation. 
There are 12 NWK commands, such as \textit{Link Status, Route Record, Route Reply}, etc. 
Zigbee's Application layer comprises of Application Support (APS) sublayer, Zigbee Device Object (ZDO), and Application Framework. 
APS sublayer maintains binding tables and address mappings, and ZDO implements the device in one of the three logical roles (ZC, ZR, or ZED). 
The application framework offers pre-defined profiles (e.g., home automation, health care, etc.) and functional domains called clusters (e.g., lighting, security, etc.) for end-manufacturers to support device interoperability.
Broadly, APL commands are either functionality specific or generic (such as \textit{Read Attributes, Report Attributes} etc.). 

\subsubsection{\textbf{Security and Privacy:}} \label{privacy}
Zigbee uses 128-bit AES encryption to provide payload confidentiality and message authentication.
The standard also has the provision for integrity-protection using 128-bit AES CCM* block cipher and replay protection using a 32-bit frame counter. 
Each Zigbee device has a pre-installed global trust center link key, which is used if the manufacturer does not provide any unique link key or QR install code. 
The Network (encryption) key is randomly generated by ZC during network formation and is common to all Zigbee nodes.

\subsection{System and Threat Model}
\begin{figure}[t]
\centering
\includegraphics[scale=1.0]{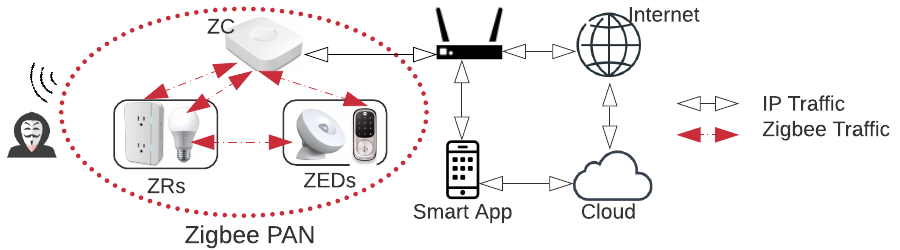}
\caption{Communication Flow in a Zigbee Home Network.}
\label{fig:zigbeepan}
\end{figure}

We assume a Zigbee home network, similar to Figure~\ref{fig:zigbeepan}, where a smart hub (ZC) is paired with several popular Zigbee devices (ZRs and ZEDs).
The hub is connected to an IP gateway to update devices' states on the cloud and the user's smart app. 
The smart home's occupants carry out routine activities and can control devices via the smart app from virtually anywhere. 
We assume that a passive attacker is collecting Zigbee transmissions using a wireless Zigbee sniffer from within the wireless communication range of the victim network. 
We use TI CC2531 Zigbee sniffer~\cite{tisniffer}, equipped with the standard omnidirectional antenna, to receive Zigbee transmissions at a distance of 20 m\footnote{Range can be extended with a high gain directional antenna}. 
The attacker does not need access to the smart app or physical presence inside the smart home; he can even implant a Zigbee sniffer nearby and observe the traffic remotely. 
 
The attacker analyses the captured Zigbee traffic to passively identify the events and devices using either command inference or periodic reporting patterns, irrespective of the network or link keys, device's QR code, or specific events like device pairing or rejoining, which aid the attacker to extract the Network key. 
In other words, we assume a fully operational Zigbee network with subject devices (door locks, bulbs, outlets, and various types of sensors) configured and commissioned a-priori.
The attacker only requires some background knowledge of the Zigbee standard. 
There is no need to collect event signatures for each device.  
The attacker needs the device's reporting signatures only when a specific device needs to be identified in the target home with zero user activity.

\subsubsection{\textbf{Challenges:}} \label{challenges}
The AES-128 algorithm used by Zigbee has proven confusion and diffusion properties and prevents eavesdropping. 
The attacker can resort to using the existing NWK command inference scheme~\cite{zigator} based on payload size, radius, and actively determined logical device type to infer NWK frames.  
Unfortunately, the events and device information is embedded in APL commands where the radius is insignificant. 
Also, unlike the 12 NWK commands, which have defined payload lengths~\cite{standard}, there are more than a hundred APL commands, most of which are manufacturer configurable (e.g., Report Attributes, Read Attributes, etc.). Hence there exist several overlappings at each payload length.  
These factors make the existing approach~\cite{zigator} insufficient to \textit{passively} infer APL commands.

The unencrypted IEEE 802.15.4 frames in the Zigbee traffic also provide negligible information regarding the devices and events, e.g., the frequently exchanged IEEE 802.15.4 ACK does not mention network or MAC address for the source or destination, and the incremental frame sequence numbers roll back after 256, making it extremely challenging to trace the communicating nodes. 

Moreover, existing research studies rely on a-priori event signatures for the identification of events~\cite{peekaboo,homonit}. 
In practice, user events are infrequent, e.g., during nighttime.  
In this idle state, the devices and hub exchange periodic reports only and do not leak any device information.   
Hence, identifying a device without event signatures or in the absence of events are still open problems for the attacker.
\section{Passive Inference Attacks on Zigbee}\label{main}
\subsection{Attack Overview}\label{sub:attack}
\begin{figure}[t]
\centering
\includegraphics[scale=0.75]{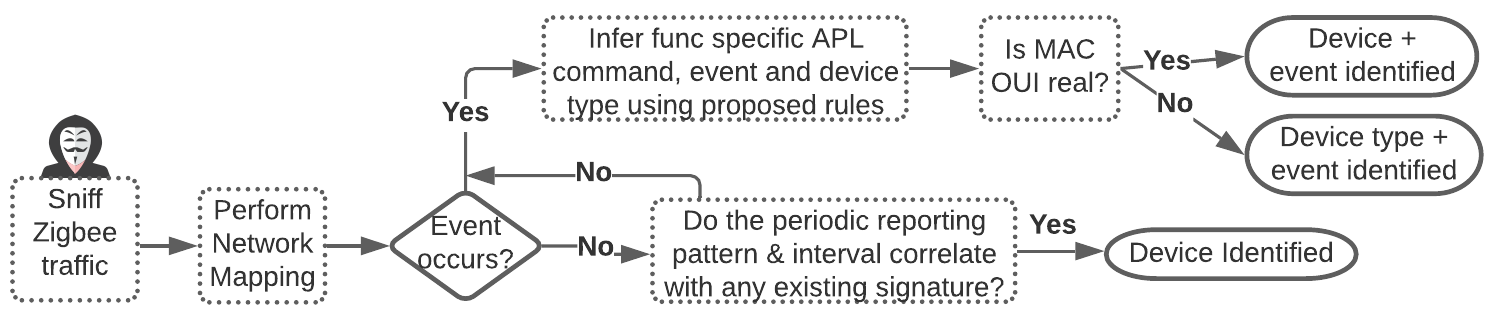}
\caption{Inference Strategy: If event occurs, infer functionality specific APL command and combine MAC identifier to identify the device and event. If there's no event, identify device using periodic signature correlation. If it fails, wait for an event.}
\label{fig:strategy}
\end{figure}

As illustrated in Figure~\ref{fig:strategy}, our fundamental goal is to invade the smart home's privacy by determining Zigbee devices, triggered events, and encrypted commands exchanged in the home. 
We use a low-cost wireless Zigbee receiver, TI-CC2531~\cite{tisniffer}, to identify and tune to the target network's communication frequency channel and sniff the Zigbee traffic.
To maximize the amount of information extracted from the sniffed traffic, we first perform network mapping, whereby the logical device type of each node (ZC, ZR, or ZED) is determined.

If an event occurs, we use proposed inference rules (Section~\ref{rules1}) to identify the functionality-specific APL command in the event's traffic, which further reveals event and device type. 
The manufacturer is revealed from the device's MAC identifier.
Specifically, we exploit the device's logical type and metadata variations in APL commands, that stem from power consumption optimizations incorporated into Zigbee.
Unlike prior works~\cite{peekaboo,homonit}, we do not require a-priori event signatures for every device and can infer unknown devices and events.

In addition, we leverage the device's reporting pattern and interval to create unique reporting signatures (Section~\ref{rules2}).
Whenever a known device with unpatched vulnerability needs to be identified in the target network with no event triggers, we correlate the device's reporting signature with the reporting pattern and interval of every device in the target's Zigbee traffic.  
If the reporting signatures are unavailable, we wait for an event to identify the device using command inference.
To the best of our knowledge, no prior work has demonstrated device identification, using APL commands, without collecting event signatures or through periodic reporting patterns. 
Below we explain the attack phases. 

\subsection{Passive Network Mapping}\label{sub:mapping}
To keep the frame length small, Zigbee uses logical address for routing, the source's MAC address for authentication, and excludes the destination's MAC address.
Thus, to identify the type and model of the target device, it is essential to keep a mapping of logical address, MAC address, and logical type (ZC, ZED, or ZR) for each logical address (i.e., node) in the traffic.
Zigbee specification~\cite{standard} identifies ZC as the node having 0x0000 logical address.
We observed that for IEEE 802.15.4 Data Requests specifically, the source node is ZED and the destination node (other than 0x0000) is ZR.
In addition, we recognized ZRs as the destination node of any Zigbee frame that has source routing information in the metadata, and that node does not send IEEE 802.15.4 Data Requests.
ZR can also be identified as the source of NWK commands namely \textit{Link Status, Rejoin Response, and Network Report}, provided the node address is not 0x0000~\cite{zigator}. 

\subsection{Device and Event Identification using Inferred APL Command}\label{sub:event1}
Although devices exhibit unique event patterns, the event traffic of same functional devices always includes same functionality-specific APL command, e.g., bulbs use \textit{color control} command for color change.
It happens because device manufacturers use defined Zigbee clusters to support vendor interoperability.
This is validated from the official Zigbee compliance documents, e.g., Lightify~\cite{osram} and Sengled~\cite{sengled} bulbs use same APL commands.  
Below we describe our scheme to devise and use command inference rules to identify events and devices.

\subsubsection{\textbf{Inference Algorithm:}}\label{rules1}
The functionality-specific APL commands of interest (\textit{OnOff, Color Control, Level Control, Lock/Unlock,} and \textit{Zone Status} (short for \textit{IAS Zone Status Change}) have fixed payload lengths. 
However, there exist overlappings with several generic APL commands within the encrypted traffic.  
This happens because there are more than hundred APL commands, many of which are manufacturer configurable and only have minimal payload and attribute size specified in the standard~\cite{zcl}. 
Thus, command \texttt{xyz} with a minimum 10-byte payload and 3-byte attribute size has a payload subset of 10, 13, 16 bytes etc. 
\begin{figure}[t]
\centering
\includegraphics[scale=0.73]{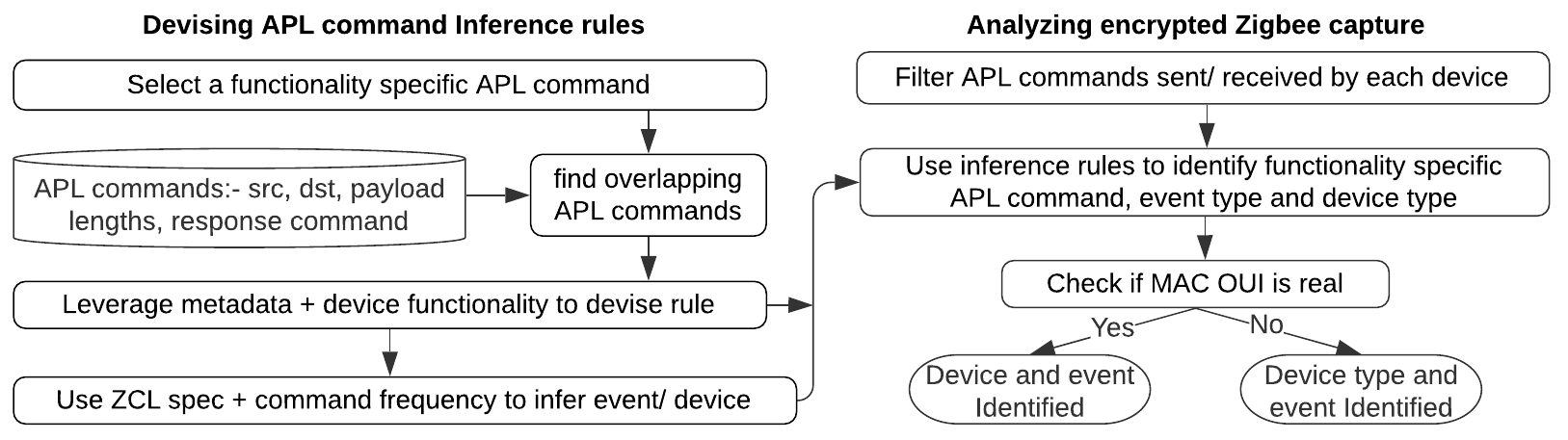}
\caption{Strategy to identify devices and events from inferred APL commands.}
\label{fig:inferenceflow}
\end{figure}

\begin{table}[t]
\centering
\caption{Identifying devices and events from Inferred commands. (Resp= Response, D=ZED/ZR, ND=NWK discovery, *=burst repeats, **=broadcast, (x)= payload len)}
\label{tab:table_genrules}
\begin{tabular}{l|l|l|l|l} 
\hline
\multicolumn{1}{c|}{\textbf{Target}}                 & \multicolumn{1}{c|}{\textbf{Inference Rule}}                                                                                                     & \multicolumn{1}{c|}{\textbf{Command}} & \multicolumn{1}{c|}{\textbf{Device Type}} & \multicolumn{1}{c}{\textbf{Event}}  \\ 
\hline
\begin{tabular}[c]{@{}l@{}}ZC-ZED(11)\\\end{tabular} & Resp = (12 \textbar{}\textbar{} 21)                                                                                                              & Lock/unlock                           & Door lock                                 & Lock/unlock                         \\ 
\hline
ZC - D(11)                                           & Resp=13\textbar{}\textbar{}15  != 12**                                                                                                           & OnOff                                 & Outlet/ bulb                              & On/ off                             \\ 
\hline
ZC - D(14)                                           & \begin{tabular}[c]{@{}l@{}}ND = 1 Resp != 11\\ Prec != 17**\end{tabular}                                                                         & Level Control                         & \multirow{2}{*}{Smart Bulb}               & Level changed                       \\ 
\cline{1-3}\cline{5-5}
ZC - D(15)                                           & ND = 1 Resp != 12                                                                                                                                & Color Control                         &                                           & Color Changed                       \\ 
\hline
\multirow{4}{*}{ZED-ZC(17)}                          & \multirow{4}{*}{\begin{tabular}[c]{@{}l@{}}Preceding Packet \textcolor[rgb]{0.502,0,0}{}\\\textcolor[rgb]{0.502,0,0}{(}Prec) != 13\end{tabular}} & Zone Status (1*)                      & Motion Sensor                             & Motion                              \\ 
\cline{3-5}
                                                     &                                                                                                                                                  & Zone Status (1)                       & Door Sensor                               & Open/ close                         \\ 
\cline{3-5}
                                                     &                                                                                                                                                  & Zone Status (2)                       & Flood sensor                              & Water leakage                       \\ 
\cline{3-5}
                                                     &                                                                                                                                                  & Zone Status (3)                       & Audio sensor                              & Audio detected                      \\
\hline
\end{tabular}
\end{table} 
As shown in Figure~\ref{fig:inferenceflow}, to devise inference rule for a functionality-specific APL command, we utilize APL frame formats~\cite{zcl} to first identify all APL commands that have overlapping payload lengths and packet direction with the target command and its response command (if any), e.g., \textit{Door Lock/ unlock request} and \textit{response}.  
Next, a test event is triggered, and overlapping commands are differentiated based on the logical device type and metadata variations (e.g., network discovery, end device initiator, etc.). 
As seen in Table~\ref{tab:table_genrules}, a command inference rule specifies properties of APL commands that must be present in the event burst (traffic). For instance, an APL command of payload length 11 bytes, sent from ZC to ZED is \textit{Lock/ unlock} command if the response packet (ZED to ZC) is 12 or 21 bytes.   
Since same functional devices use same functionality-specific commands, the inference rules constructed for a certain device also hold true for other manufacturers' devices.  
We stress-checked the rules against 200 MBs of Zigbee capture from our devices and third-party sources~\cite{crawdad,tshark}.
Note that most APL commands (like \textit{color control}), directly reflect the event and device type.
However, for outlets and bulbs, that use same \textit{OnOff} command, the device type is indistinguishable until an additional event, e.g., color change is triggered.
For \textit{Zone Status} command, we observed behavioral consistencies that allowed us to differentiate various types of sensors; e.g., \textit{Zone Status} appears twice in the event burst for flood sensor and thrice for the audio sensor. 
For motion and door sensor, \textit{Zone Status} appears once only. However, we noticed that for motion sensors, the same burst pattern repeats after few seconds.

\subsubsection{\textbf{Identifying Events and Devices:}}
We first filter all APL commands in the event's traffic sent to or received by the target logical address (e.g., 0xabcd) and discard any duplicate packet.
We observed that the functionality-specific command is generally the first APL command of the event burst; hence we also discard bursts that do not have any frame with target payload lengths (11-17 bytes) in the initial half of the burst. 
If a command with target payload length exists, we use Table~\ref{tab:table_genrules} to identify the APL command, event, and device type. 
Finally, we combine the manufacturer's identity extracted from the device's MAC OUI (e.g., PhilipL) to identify the device. 
Note that the exact device's identification depends on the MAC OUI showing real manufacturer, rather than system-on-chip (SOC) manufacturer, e.g., SiliconL.
In essence, we can passively identify unknown events and devices from the target functional domains (bulbs, outlets, door locks, and sensors) without the Network key or event signatures.
\subsection{Device Identification using Periodic Reporting Patterns}\label{sub:event2}
Zigbee devices periodically report their status (battery level, firmware upgrades, etc.) to ZC. 
Since every functional device has varied power consumption, the manufacturers manipulate periodic reporting frequency, and specific frame attributes to comply with the Zigbee certification requirement of minimum 2 years battery life~\cite{batteryreq}.
The discrepancies in reporting patterns and intervals allow us to devise unique device fingerprints and identify devices even when no event occurs (e.g., during office hours).
Unlike event bursts, reporting bursts have no functionality-specific APL command and do not directly reveal device's identity. 

\begin{figure}[t]
\centering
\includegraphics[scale=0.65]{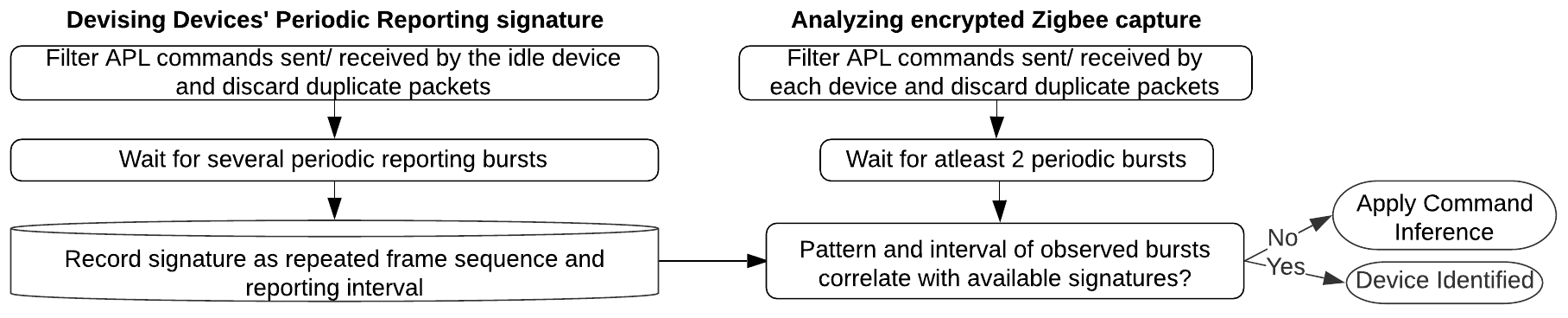}
\caption{Strategy to identify devices using periodic reporting patterns.}
\label{fig:periodicflow}
\end{figure}

\subsubsection{\textbf{Devising Periodic Reporting Signatures: }}\label{rules2}
As shown in Figure~\ref{fig:periodicflow}, to devise a device's periodic reporting signature, we put the device in the idle state and filter all APL commands exchanged between the device and ZC. 
After discarding duplicate packets, we analyze the traffic to determine at least three bursts with same reporting pattern and interval. 
Thus, the signature sign\textsubscript{i} is a sequence of APL frames f\textsubscript{i} defined using logical device type of source (src) and destination (dst), payload length (pl), and reporting interval (RI) and is represented as:
\begin{equation}
sign_i = \{f_1 , f_2 , f_3, ...\}  \mbox{  where   } f_i = \{src_i,  dst_i,  pl_i,  RI_i\} 
\end{equation} 

\subsubsection{\textbf{Identifying Devices:}} 
We first filter APL commands from the traffic and discard duplicate packets or bursts with any functionality-specific command. 
Next, we look for two similar bursts and correlate the observed pattern and interval with the available signature set to identify the device. 
Rarely, but if two signature sets collide, we use additional attributes like MAC OUI to make a decision.
If no reporting signatures are available for a device, we wait for an event burst to identify the device using command inference approach (Section~\ref{rules1}).
\section{Experimental Setup and Results}
\subsection{Automating Passive Inference Attacks with ZLeaks Tool}\label{sub:zleaks}
To automate the inference attacks depicted in Figure~\ref{fig:strategy}, we developed a command-line tool in Python, named ZLeaks. 
ZLeaks takes Zigbee PCAP capture as input and determines the event occurrences and devices in the network.   
While in the vicinity of the target network, the attacker can run ZLeaks on his laptop or embedded board like Raspberry Pi with a single command.
ZLeaks extracts all APL commands from the captured traffic and uses Pyshark library~\cite{pyshark} to parse required frame attributes (e.g., payload length, logical types of nodes, etc.) in a temporary CSV file for analysis.
ZLeaks then attempts to identify events and devices using either proposed APL inference rules (Section~\ref{sub:event1}) or available reporting signatures (Section~\ref{sub:event2}).
Note that the attacker can automatically extract reporting signatures of an idle Zigbee device using ZLeaks Signature Extractor.  

\subsection{Experimental Setup}
Our device set comprised of 27 commercial off-the-shelf Zigbee devices (ranging from bulbs, locks, outlets, to various sensors) that were selected based on Amazon's popularity and manufacturer diversity. 
Amongst 27 devices, 19 devices were unique, while a few non-unique devices were purchased from a different source and tested to ensure that the evaluation results for a particular device and model remain consistent.   
Furthermore, while we used 11 unique devices to formulate inference strategy, we set aside 8 unique devices, at least one from each functional domain as the unknown devices for the sole purpose of evaluation. The known and unknown devices are listed for reference in Table~\ref{tab:controlled1} and Table~\ref{tab:realistic1} respectively.
The tests were conducted with 2 universal (manufacturer-independent) hubs; SmartThings and Amazon Echo Plus and 3 vendor-specific hubs; Philips Hue Bridge 2.1, Sengled Z02-hub, and Lightify Gateway. 
This is by far the most extensive Zigbee device set used to evaluate Zigbee protocol. 

We evaluated ZLeaks identification techniques in following three settings;

\begin{figure}[t]
\centering
\includegraphics[scale=1.0]{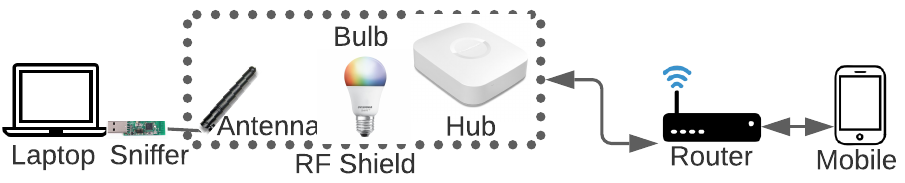}
\caption{Experimental Setup for analyzing Zigbee Devices: SMA cable connects sniffer's antenna (inside RF shield) with TI CC2531 sniffer connected to the laptop outside.}
\label{fig:experiment}
\end{figure}

\textbf{RF shield:}
It was used to i) study devices' response to event triggers while devising command inference rules, ii) collect the device's reporting pattern, and iii) perform a controlled evaluation of ZLeaks by simultaneously pairing multiple devices with each hub.  
As depicted in Figure~\ref{fig:experiment}; the RF shield was connected to the gateway to provide continued Internet access to ZC placed inside the shield. 
To sniff the Zigbee communication, a standard omnidirectional antenna (inside the shield) was connected via an SMA cable to a low-cost TI CC2531 wireless Zigbee sniffer~\cite{tisniffer} plugged into the laptop (outside the shield). 

\textbf{IoT ``Living Lab":}
It is a realistic noisy IoT lab named \enquote{Mon(IoT)r Lab} at Northeastern University~\cite{moniotr}, which has more than 100 smart devices already connected over several wireless networks, along with various non-IoT networks. 

\textbf{Public Captures:}
We used Zigbee captures from; i) Wireshark forum~\cite{wshark1}, and ii) Prior captures~\cite{zigator} available on Crawdad~\cite{crawdad} to show that ZLeaks is independent of evaluation testbed, device set, and works for unknown devices.
We verified the results using the Network keys available with the capture files.
Both the captures contained only event bursts and did not include enough reporting patterns to evaluate the periodic reporting approach.

\subsection{Evaluation Metrics}
We evaluated ZLeaks using three parameters; 1) Inferred APL commands, 2) Event and device type extracted from APL command, and 3) Correlated periodic reporting patterns. 
We used traditional accuracy metrics to evaluate parameters 1 and 3. 
As a particular inferred APL command always yields same results for event and device, we evaluated parameter 2 using proposed \textit{Device Score} scheme.
 
\textbf{Traditional Metrics:}
We use True Positive Rate (TPR) and False Negative Rate (FNR) to specify the rate of correct and missed (or out-of-order) observations, respectively. 
As evaluation results indicate, there are no False Positives (FP) or True Negative (TN) outcomes, hence, we calculate \textit{accuracy}, i.e., the ratio of correctly inferred observations to the total number of observations, as:   
\begin{equation}
\mbox{TPR (recall)} = \frac{\mbox{TP}}{\mbox{TP + FN'}}
\end{equation}
\begin{equation}
\mbox{FNR} = \frac{\mbox{FN}}{\mbox{TP + FN'}}
\end{equation}
\begin{equation}
\mbox{Accuracy} = \frac{\mbox{TP+TN}}{\mbox{TP + TN + FP + FN'}} 
\end{equation}

\textbf{Score (\textit{short for Device Score}): }
It determines the amount of device and event information extracted from the inferred APL command and device OUI. 
We calculate Score as a sum of device type (DT), event type (ET) and manufacturer's identity (M), with weights of each attribute defined in Table~\ref{tab:score}.
\begin{equation}
\mbox{Score} = M + DT + ET
\end{equation}

To understand Score, consider switching on a bulb that triggers a functionality-specific APL command from ZC to ZED of payload size 11 bytes.
The highest Score is 5 when all attributes are correctly inferred, and lowest is 0 when nothing is inferred. 
As per Table~\ref{tab:table_genrules}, the command is either \textit{Lock/ unlock or On/off}. 
From Table~\ref{tab:score}, DT and ET are 1 if these two commands are indistinguishable. For \textit{On/off} command, DT (bulbs or outlet) and ET (on or off) are 1.5, whereas for \textit{Lock/ unlock} command, DT is 2 (lock) while ET is 1.5 (lock or unlock).  

\begin{table}[t]
\centering
\caption{Score Table for Evaluating Command Inference Approach}
\label{tab:score}
\begin{tabular}{c|l|c} 
\hline
\textbf{Attributes}               & \textbf{Score}   & \textbf{Example}                             \\ 
\hline
\multirow{2}{*}{Manufacturer (M)} & 0 = SOC OUI      & SiliconL, Ember, TexasIns, NordiacSE ..      \\ 
\cline{2-3}
                                  & 1 = Real MAC OUI & PhilipsL, OSRAM, SmartThi, Zhejiang ..       \\ 
\hline
\multirow{4}{*}{Device Type (DT)} & 0 = Unidentified & -                                            \\ 
\cline{2-3}
                                  & 1 = Uncertain    & door lock or bulb (different commands)       \\ 
\cline{2-3}
                                  & 1.5 = Indistinct & either outlet or bulb? (same command)        \\ 
\cline{2-3}
                                  & 2 = Identified   & Outlet, door lock, motion sensor, bulb ..    \\ 
\hline
\multirow{4}{*}{Event Type (ET)}  & 0 = Unidentified & -                                            \\ 
\cline{2-3}
                                  & 1 = Uncertain    & lock/unlock or on/off (different commands)   \\ 
\cline{2-3}
                                  & 1.5 = Indistinct & either door lock or unlock? (same commands)  \\ 
\cline{2-3}
                                  & 2 = Identified   & motion detected, color change, etc ..        \\
\hline
\end{tabular}
\end{table}
\begin{table}[t]
\centering
\caption{Controlled Evaluation: Identifying Devices and Events using Inferred APL Commands. Here, SMT= SmartThings, M= Manufacturer, DT= Device Type, ET= Event type, and *= burst repeats after few seconds}
\label{tab:controlled1}
\begin{tabular}{l|l|l|l|c|c|c|c} 
\hline
\textbf{Device (Model)}                         & \textbf{Event } & \textbf{OUI} & \textbf{Command (\#)} & \textbf{M} & \textbf{DT} & \textbf{ET} & \textbf{Score}  \\ 
\hline
\multirow{4}{*}{\begin{tabular}[c]{@{}l@{}}Philips Hue Color Bulb \\(LCA-003)\end{tabular}} & Off                                                                                              & PhilipsL                                                                                        & Off with effect                                                                                        & 1                               & 2                                & 2                                & 5                                   \\ 
\cline{2-8}
                                        & On                                                                                               & PhilipsL                                                                                        & On/off: On                                                                                             & 1                               & 2                                & 2                                & 5                                   \\ 
\cline{2-8}
                                        & Color change                                                                                     & PhilipsL                                                                                        & Color Control                                                                                          & 1                               & 2                                & 2                                & 5                                   \\ 
\cline{2-8}
                                        & Dim                                                                                              & PhilipsL                                                                                        & Level Control                                                                                          & 1                               & 2                                & 2                                & 5                                   \\ 
\hline
\multirow{3}{*}{\begin{tabular}[c]{@{}l@{}}Sengled Color Bulb \\(E11-N1EA)\end{tabular}}   & Color change                                                                                     & Zhejiang                                                                                        & Color Control                                                                                          & 1                               & 2                                & 2                                & 5                                   \\ 
\cline{2-8}
                                        & Dim                                                                                              & Zhejiang                                                                                        & Level Control                                                                                          & 1                               & 2                                & 2                                & 5                                   \\ 
\cline{2-8}
                                        & On/off                                                                                           & Zhejiang                                                                                        & On/Off                                                                                                 & 1                               & 1.5                              & 1.5                              & 4                                   \\ 
\hline
Sengled White Bulb G14                     & On/Off                                                                                           & Zhejiang                                                                                        & On/Off                                                                                                 & 1                               & 1.5                              & 1.5                              & 4                                   \\ 
\hline
Centralite Outlet (Mini)                       & On/Off                                                                                           & siliconL                                                                                        & On/Off                                                                                                 & 0                               & 1.5                              & 1.5                              & 3                                   \\ 
\hline
Sonoff Outlet  (S31 Lite)                          & On/Off                                                                                           & texasIns                                                                                        & On/Off                                                                                                 & 0                               & 1.5                              & 1.5                              & 3                                   \\ 
\hline
SMT Outlet (US-2)                             & On/Off                                                                                           & Smartthi                                                                                        & On/Off                                                                                                 & 1                               & 1.5                              & 1.5                              & 4                                   \\ 
\hline
SMT Motion sensor (IM)                 & Motion                                                                                           & Smartthi                                                                                        & Zone Status (1*)                                                                                       & 1                               & 2                                & 2                                & 5                                   \\ 
\hline
SMT Multisensor (250)                        & Open/close                                                                                       & samjin                                                                                          & Zone Status (1)                                                                                        & 1                               & 2                                & 1.5                              & 4.5                                 \\ 
\hline
Ecolink Water Sensor                     & water leak                                                                                       & ember                                                                                           & Zone Status (2)                                                                                        & 0                               & 2                                & 2                                & 4                                   \\ 
\hline
Ecolink Sound Sensor                    & Sound                                                                                            & ember                                                                                           & Zone Status (3)                                                                                        & 0                               & 2                                & 2                                & 4                                   \\ 
\hline
Yale Door lock (YRD226)                         & Lock/unlock                                                                                      & ember                                                                                           & Lock/Unlock                                                                                            & 0                               & 2                                & 1.5                              & 3.5                                 \\
\hline
\end{tabular}
\end{table}
\begin{table}[h!]
\centering
\caption{Realistic Evaluation: Identifying Unknown Devices and Events using Inferred Commands. (M= Manufacturer, DT= Device Type, ET= Event type, *= repeats)}
\label{tab:realistic1}
\begin{tabular}{l|l|l|l|c|c|c|c} 
\hline
\textbf{Device (Model)}                         & \textbf{Event } & \textbf{OUI} & \textbf{Command (\#)} & \textbf{M} & \textbf{DT} & \textbf{ET} & \textbf{Score}  \\ 
\hline
\multirow{2}{*}{Philips White Bulb}  & Off                                                                                             & \multirow{2}{*}{PhilipsL}                                                                & Off with effect                                                                                        & 1          & 2           & 2           & 5               \\ 
\cline{2-2}\cline{4-8}
                                     & On                                                                                              &                                                                                          & On/off: On                                                                                             & 1          & 2           & 2           & 5               \\ 
\hline
\multirow{3}{*}{\begin{tabular}[c]{@{}l@{}}OSRAM Color Bulb \\(Sylvania Smart+)\end{tabular}}    & On/off                                                                                          & ledvance                                                                                 & On/Off                                                                                                 & 1          & 2           & 1           & 4               \\ 
\cline{2-8}
                                     & Color change                                                                                    & ledvance                                                                                 & Color Control                                                                                          & 1          & 2           & 2           & 5               \\ 
\cline{2-8}
                                     & Dim                                                                                             & ledvance                                                                                 & Level Control                                                                                          & 1          & 2           & 2           & 5               \\ 
\hline
SmartThings (SMT) Bulb                     & On/off                                                                                          & SiliconL                                                                                 & On/Off                                                                                                 & 0          & 1.5         & 1.5         & 3               \\ 
\hline
Aqara Outlet (US)                        & On/off                                                                                          & jennic                                                                                   & On/Off                                                                                                 & 1          & 1.5         & 1.5         & 4               \\ 
\hline
Ewelink Outlet  (SA-003)                     & On/off                                                                                          & TexasIns                                                                                 & On/Off                                                                                                 & 0          & 1.5         & 1.5         & 3               \\ 
\hline
SMT Motion Sensor IRM               & Motion                                                                                          & samjin                                                                                   & Zone Status (1*)                                                                                       & 1          & 2           & 2           & 5               \\ 
\hline
Visonic Door sensor MCT                 & Open/close                                                                                      & ember                                                                                    & Zone Status (1)                                                                                        & 0          & 2           & 1.5         & 3.5             \\ 
\hline
Schlage Lock (Connect)                 & Lock/unlock                                                                                     & siliconL                                                                                 & Lock/Unlock                                                                                            & 0          & 2           & 1.5         & 3.5             \\
\hline
\end{tabular}
\end{table} 
\subsection{Device and Event Identification using Inferred APL Command}
\subsubsection{\textbf{Controlled Evaluation in RF Shield:}}\label{controlled1}
We simultaneously paired all compatible devices with one hub at a time inside the RF shield and generated events randomly. 
From the sniffed traffic, ZLeaks inferred functionality-specific APL commands and MAC OUI for each device to determine triggered events and devices. 
Since the inferred APL command and MAC OUI remain same for a particular device-event pair (e.g., color change for Sengled bulb), the Score remains same for every event prompt irrespective of the hub. 
Therefore, Table \ref{tab:controlled1} reports findings of each device once.
We see that distinct events like color change, motion detected, etc., are easy to infer than binary events (e.g., on/off). 
Philips bulb is an exception here as it uses distinct commands to represent on and off events. 
Furthermore, we identified various sensors from a single \textit{Zone Status} command based on behavioral consistencies (refer to rules in Table~\ref{tab:table_genrules}).  
To conclude, the Score is dependent on the correct identification of the APL command and the MAC OUI showing the real manufacturer e.g., PhilipsL (Philips), SmartThi/ Samjin (SmartThings), Ledvance (OSRAM), Zhejiang (Sengled), Jennic (Aqara), etc.
ZLeaks identified all devices with an average Score of 4.3 out of 5 (i.e., 86.3\% information was successfully extracted). 

\subsubsection{\textbf{Realistic Evaluation in an IoT \enquote{Living Lab}:}} \label{realistic1}
Next, we shifted all these devices, hubs, and 8 unknown (unseen) devices to the IoT lab. 
Again, we simultaneously paired all devices to one hub at a time, generated random events, and analyzed the traffic with ZLeaks. 
Despite the noisy environment, the known devices exhibited the same Score as reported in Table \ref{tab:controlled1}. 
The experimental results for unknown devices are presented in Table~\ref{tab:realistic1}. 
Unknown devices with real MAC OUI and distinct event types, e.g., color change for Sengled bulb, were accurately identified by ZLeaks. 
Overall, ZLeaks identified unknown devices with an average Score of 4.2 out of 5 (i.e., identified 83.6\% devices and events). 
We conclude that despite devices exhibiting unique event signatures across different hubs, the functionality-specific APL command remains same and can be used to effectively identify any unknown device with a single event trigger.

\begin{table}[t]
\centering
\caption{Public Evaluation: Identifying Unknown Devices and Events using Inferred Commands. (M= Manufacturer, DT= Device Type, ET= Event type, *= repeats)}
\label{tab:public1}
\begin{tabular}{l|l|l|l|c|c|c|c} 
\hline
\textbf{Source}                                                              & \textbf{Unknown Device}                                               & \textbf{MAC OUI} & \textbf{Command (\#)} & \textbf{M} & \textbf{DT} & \textbf{ET} & \textbf{Score}  \\ 
\hline
\multirow{2}{*}{\begin{tabular}[c]{@{}l@{}}Wireshark \\ZCL log~\cite{wshark1}\end{tabular}} & Motion Sensor 1                                                       & private          & Zone Status (1*)      & 0          & 2           & 2           & 4               \\ 
\cline{2-8}
                                                                             & Motion Sensor 2                                                       & none             & Zone Status (1*)      & 0          & 2           & 2           & 4               \\ 
\hline
\begin{tabular}[c]{@{}l@{}}Zigator~\cite{crawdad} \\Sth2-duos\end{tabular}                  & \begin{tabular}[c]{@{}l@{}}SmartThings Outlet\\(IM6001)\end{tabular} & samjin           & On/off                & 1          & 1.5         & 1.5         & 4               \\
\hline
\end{tabular}
\end{table} 
\subsubsection{\textbf{Open World Evaluation on Public Captures:}} \label{public1}
We evaluated ZLeaks over public Zigbee captures and reported results in Table~\ref{tab:public1}. 
In capture 1~\cite{wshark1}, we found 2 unknown devices that were recognized as motion sensors due to the presence of repetitive \textit{Zone Change} commands. 
Capture 2~\cite{crawdad} had 1 unknown device which used \textit{On/Off} for events. 
Note that we removed device commissioning traffic (including Network key) from both files to comply with our threat model.   

As device identification is dependent on the correct inference of functionality-specific APL commands, we also evaluated ZLeaks inference rules on capture 2~\cite{crawdad}.
The results in Table~\ref{tab:apl_cmds} indicate that ZLeaks inferred functionality-specific APL commands with 91.2\% accuracy.
We used our command inference strategy on generic APL commands and were able to infer \textit{Device Announcement, Bind Request and Response (RR), Link Quality RR, NWK Address RR, Parent Announcement RR,} etc., with 100\% accuracy. 
Most of all, the 6 NWK commands that Zigator~\cite{zigator} could not identify, were inferred with 85.7\% accuracy.
\begin{table}[t]
\centering
\caption{Evaluating APL Command Inference rules on Public Zigbee Capture~\cite{crawdad}. Note: * implies that the command is identified, but not the state.}
\label{tab:apl_cmds}
\begin{tabular}{l|c|c|c} 
\hline
\multicolumn{1}{c|}{\textbf{APL Commands}}     & \textbf{Total Packets} & \multicolumn{1}{l|}{\textbf{Inferred Packets}} & \textbf{Accuracy (\%)}  \\ 
\hline
Zone Status Change                             & 2916                   & 2712                                           & 93.0                    \\ 
\hline
ZCL On \textbar{}\textbar{} ZCL Off         & 2423                   & 2175*                                          & 89.8                    \\ 
\hline
Door lock \textbar{}\textbar{} Unlock Request  & 676                    & 596*                                           & 88.1                    \\ 
\hline
Door lock \textbar{}\textbar{} Unlock Response & 403                    & 370 *                                          & 91.8                    \\ 
\hline
Color Control, Level Control                   & 0                      & 0                                              & 0                       \\
\hline
\end{tabular}
\end{table}
\subsection{Device Identification using Periodic Reporting Patterns}
\subsubsection{\textbf{Controlled Evaluation in RF Shield:}} \label{controlled2}
We simultaneously paired all known devices with one hub at a time in an RF shield and left them in the idle state for at least 3 hrs.
This way, devices reporting the attributes every 5 or 10 mins yielded 36 and 18 reporting patterns, respectively, which are sufficient to evaluate two main features; reproducibility and uniqueness of periodic signatures.
Table \ref{tab:controlled2} summarizes the results of this experiment, with reporting intervals in second, minute, and hour represented using letters s, m, and h.
Note that several devices exhibited more than one reporting pattern, e.g., for battery, temperature, etc., while few devices showed a different number of reporting patterns across different hubs, e.g., SMT and Sonoff outlet. 
This essentially helped identify both the device and the smart hub from the encrypted traffic.
It is evident from a high average TPR of 0.998 and low FNR of 0.002 that the periodic signatures were identifiable and consistent over time, except once when the Centralite outlet and SMT Multisensor showed two out-of-order packets and were not identified.

\begin{table}[t]
\centering
\caption{Controlled Evaluation of Periodic Reporting scheme. Here, SMT= SmartThings, RI= Reporting interval, TPR= True positive rate, FNR = False negative rate}
\label{tab:controlled2}
\begin{tabular}{l|c|c|c|c|c|c|c|c|c} 
\hline
\multicolumn{1}{c|}{\multirow{2}{*}{\textbf{Device }}} & \multicolumn{3}{c|}{\textbf{SMT v2 Hub}} & \multicolumn{3}{c|}{\textbf{Amazon Echo+}} & \multicolumn{3}{c}{\textbf{Philips/ Sengled}}  \\ 
\cline{2-10}
\multicolumn{1}{c|}{}                                  & RI      & TPR   & FNR                    & RI      & TPR   & FNR                      & RI       & TPR & FNR                            \\ 
\hline
Centralite Outlet                                      & 5, 10m  & 1     & 0                      & 5, 9m   & 0.982 & 0.018                    & \multicolumn{3}{c}{N/A}                        \\ 
\hline
Sonoff Outlet                                          & 5m      & 1     & 0                      & 5, 10m  & 1     & 0                        & \multicolumn{3}{c}{N/A}                        \\ 
\hline
SMT Outlet                                             & 5, 10m  & 1     & 0                      & 10m     & 1     & 0                        & \multicolumn{3}{c}{N/A}                        \\ 
\hline
Sengled White Bulb                                     & 5m      & 1     & 0                      & 10m     & 1     & 0                        & 5,20,25m & 1   & 0                             \\ 
\hline
Sengled Color Bulb                                     & 10m, 1h & 1     & 0                      & 10m, 1h & 1     & 0                        & 5,20,25m & 1   & 0                             \\ 
\hline
Philips Hue Color Bulb                                 & 1s, 2m  & 1     & 0                      & 1s, 2m  & 1     & 0                        & 1s, 2m   & 1   & 0                             \\ 
\hline
SMT Motion sensor (IM)                                   & 5m      & 1     & 0                      & 5m      & 1     & 0                        & \multicolumn{3}{c}{N/A}                        \\ 
\hline
SMT Multisensor                                        & 5m, 1h & 0.975 & 0.025                  & 5m, 1h  & 1     & 0                        & \multicolumn{3}{c}{N/A}                        \\ 
\hline
Ecolink water sensor                                   & 30, 30m & 1     & 0                      & \multicolumn{3}{c|}{N/A}                   & \multicolumn{3}{c}{N/A}                        \\ 
\hline
Ecolink sound sensor                                   & 27, 30m & 1     & 0                      & \multicolumn{3}{c|}{N/A}                   & \multicolumn{3}{c}{N/A}                        \\ 
\hline
Yale Door Lock                                         & 1h      & 1     & 0                      & 10m     & 1     & 0                        & \multicolumn{3}{c}{N/A}                        \\
\hline
\end{tabular}
\end{table}
\begin{table}[t]
        \centering
        \caption{Realistic Evaluation of Periodic Reporting scheme. (\cf  = successful device identification, \ch  = success using additional info, and \ce  = resembled other device)}
        \label{tab:realistic2}
          \begin{tabular}{l|c|c|c|c|c|c|c|c|c|c|c|c} 
                        \hline
\multicolumn{1}{c|}{\multirow{2}{*}{\textbf{Device}}} & \multicolumn{4}{c|}{\textbf{SMT v2 Hub}} & \multicolumn{4}{c|}{\textbf{Amazon Echo+}} & \multicolumn{4}{c}{\textbf{Vendor Hub}}  \\ 
\cline{2-13}
\multicolumn{1}{c|}{}                                  & 15m        & 30m      & 1h & 3h    & 15m    & 30m & 1h & 3h           & 15m                   & 30m & 1h & 3h   \\ 
\hline

Centralite Outlet                                       & \cf         & \cf       & \cf  & \cf     & \cf     & \cf  & \cf  & \cf             &                       &     &     &       \\ 
\hline
Sonoff Outlet                                           & \cf         & \cf       & \cf  & \cf     & \cf     & \cf  & \cf  & \cf             &                       &     &     &       \\ 
\hline
SmartThings (SMT) Outlet                                              & \cf         & \cf       & \cf  & \cf     & \cf     & \cf  & \cf  & \cf             &                       &     &     &       \\ 
\hline
Sengled (White) Bulb                                        & \cf         & \cf       & \cf  & \cf     & \cf     & \cf  & \cf  & \cf             & \cf                    & \cf  & \cf  & \cf    \\ 
\hline
Sengled (Color) Bulb                                        & \cf & \cf       & \cf  & \cf     & \cf     & \cf  & \cf  & \cf             & \cf                    & \cf  & \cf  & \cf    \\ 
\hline
Philips Hue (Color) Bulb                                    & \cf         & \cf       & \cf  & \cf     & \cf     & \cf  & \cf  & \cf             & \cf                    & \cf  & \cf  & \cf    \\ 
\hline
SMT Motion sensor (IM)                                    & \ch & \cf       & \cf  & \cf     & \ch & \cf  & \cf  & \cf             & \multicolumn{1}{c|}{} &     &     &       \\ 
\hline
SMT Multi sensor                                 & \ce  & \ce   & \cf  & \cf     & \ch & \cf  & \cf  & \cf             & \multicolumn{1}{c|}{} &     &     &       \\ 
\hline
Ecolink water sensor                                    &            & \cf       & \cf  & \cf     & \multicolumn{4}{c|}{Not compatible} &                       &     &     &       \\ 
\hline
Ecolink sound sensor                                    &            & \cf & \cf  & \cf     & \multicolumn{4}{c|}{Not compatible} &                       &     &     &       \\ 
\hline
Yale Door Lock                                          &            &          & \ch  & \ch     & \cf     & \cf  & \cf  & \cf             & \multicolumn{1}{c|}{} &     &     &       \\
\hline
        \end{tabular}
\end{table}

\subsubsection{\textbf{Realistic Evaluation in IoT \enquote{Living Lab}:}} \label{realistic2}
Next, we shifted all the hubs, and known and unknown devices to the Mon(IoT)r lab. 
We paired all compatible devices to one target hub at a time and used the remaining devices as background Zigbee noise sources. 
The devices were left in the idle state for 3 hrs, and ZLeaks analyzed traffic after specific time intervals (15 mins, 30 mins, 1 hr, and 3 hrs).
In Table \ref{tab:realistic2}, devices that were distinctively identified after the specified time are marked with a full circle, e.g., all outlets had reporting intervals of 5 and/or 10 mins; and were successfully identified within 15 mins.
Half-circle indicates identical reporting pattern and interval for two devices, e.g., both Yale and Schlage lock (unknown device) reported the same pattern after 1 hr. 
In such a case, we used additional parameters (e.g., MAC OUI or logical device type) to identify the device.
Finally, an empty circle depicts a complete resemblance between two devices, e.g., SMT motion sensor IRM (an unknown device) and SMT multisensor showed same patterns until the latter device reported its second pattern after 1 hr.  
In essence, it is quite concerning that the device and manufacturer identity is leaked even in the device's idle state.
Note that we could not evaluate this approach on public captures due to the absence of periodic reporting patterns.
\section{Discussion and Related Work}
\subsection{Security Implications of Leaked Data}
The know-how of devices in the smart home and their states (e.g., door unlocked or bulb off) is crucial to the smart home's security. 
A burglar can use this information to get insight into users' affluence and determine when the house is vulnerable to intrusion. 
In addition, an attacker can use Common Vulnerabilities and Exposures (CVE) database~\cite{cve} to find and exploit unpatched vulnerabilities in the identified devices. 
The vulnerable devices can be weaponized to spread malware to the network~\cite{iotworm}, create IoT botnets~\cite{iotnet} or carry out denial of service attacks. 
The attacker may also use side-channel attack to hijack the vulnerable hub~\cite{laser}. 
From a business perspective, the leaked information can help Zigbee manufacturers gain deep insights into users' usage and activity patterns. 
This information can be sold to advertisers for interest-based advertisements, online tracking, or used in business decisions on future products. 
In short, our study provides deep insights into potential information leakages right at the source. 

\subsection{Potential Countermeasures}
ZLeaks demonstrates the significance of unencrypted metadata (MAC OUI, frame, and payload lengths) in identifying functionality specific commands, events, and devices in the Zigbee network.
Although exponential padding~\cite{padding} effectively disguises payload lengths, it adds transmission overhead and increases power consumption for low-power Zigbee devices.
We suggest padding random bytes in each payload (e.g., 0,1,2 or 3 bytes) and using the reserved field in the Zigbee security header to denote the number of padded bytes. 
This way, even same APL commands will have four different payload sizes, which will add enough entropy to make the Zigbee commands indistinguishable. 
Secondly, Zigbee Alliance can mandate the use of chipset manufacturer's identifier as MAC OUI to hide the real manufacturer's identity. 
This alone reduces the average Score calculated for unknown devices using the APL inference approach from 4.0 (80\%) to 3.1 (62\%).

The volumetric analysis also provides significant hints regarding the occurrence of an event or periodic reporting.
To make events indistinguishable, prior research~\cite{isend,sniffmislead} leverage mains-powered ZR or ZC to inject decoy packets in the traffic at pseudo-random intervals. However, decoy injector requires continuous training to avoid detection by the attacker.   
An efficient way to disguise events is to have similar event responses and reporting patterns for all devices. 
Alternatively, all attributes can be pipelined in a single packet instead of a series of packets.
The suggested countermeasures, as summarized in Table~\ref{tab:countermeasures} require significant design and implementation changes in the Zigbee protocol, as it is hard to prevent proposed inference attacks with a simple workaround like using a secure network or link key.
We believe this is why the Zigbee Alliance is involved in new smart home technology, Matter~\cite{matter}, which has security as the fundamental design tenet and does not use Zigbee as the underlying IoT protocol.

\begin{table}[t]
\centering
\caption{Privacy Implications of Leaked Information and Suggested Countermeasures.}
\label{tab:countermeasures}
\begin{tabular}{l|l|l} 
\hline
\textbf{Info}    & \textbf{Privacy Implication}                                                                                                                               & \multicolumn{1}{c}{\textbf{Countermeasure to disguise info}}                                                                \\ 
\hline
\multirow{2}{*}{Device} & \multirow{2}{*}{\begin{tabular}[c]{@{}l@{}}- Hijack vulnerable devices/ network\\- Insight regarding user's affluence\\- Online advertising\end{tabular}} & Use SOC MAC OUI to hide vendor \\ 
\cline{3-3}
                        &                                                                                                                                                            & \begin{tabular}[c]{@{}l@{}}Use random payload padding to \\complicate command inference\end{tabular}                             \\ 
\hline
Event                   & \begin{tabular}[c]{@{}l@{}}- Reveals user's presence/ absence \\ ~~daily routine\end{tabular}                                                                & \begin{tabular}[c]{@{}l@{}}Attribute Pipelining \textbf{OR} similar request\\-response patterns to hide events\end{tabular}  \\
\hline
\end{tabular}
\end{table}
\subsection{Related Work}
\textbf{Privacy Analysis of Smart Home's IP Traffic:}
Several research studies have analysed the encrypted IP network traffic of smart homes to predict devices' events~\cite{event1,pingpong}, user's habits~\cite{anybodyhome}, device types~\cite{audi,profiliot,devtype1,iothound,thangavelu19deft}, and network anomalies~\cite{cho2016fingerprinting,maldect1}. 
Few studies~\cite{homonit,app1} also analyzed the IP traffic between the smart app and cloud to detect misbehaving smart apps.
Although these studies yield promising results, there are a few limitations; 1) attacker requires physical access to the network or mobile app, and 2) these approaches exploit traffic metadata (i.e., payload length, DNS responses, etc.); hence their effectiveness is questionable under realistic network conditions like Virtual Private Networks and Network Address and Port Translation enabled.  
Although recent studies have leveraged packet-level signatures and temporal packet relations to identify events~\cite{pingpong} and devices~\cite{iotfinder} despite traffic shaping in place, these machine learning (ML) approaches require re-training after firmware upgrades to extract new signatures.

\textbf{Privacy analysis of Zigbee (non-IP) Traffic:}
Unlike IP traffic patterns, Zigbee traffic patterns are challenging to obfuscate using conventional traffic shaping, as it directly impacts power consumption and battery life.
Still, very few studies~\cite{peekaboo,homonit,ziot,iotgaze,iotspy} have analyzed Zigbee traffic with the intent to study information leakages right at the source.
Zigator~\cite{zigator} exploited unencrypted attributes of Zigbee frames, notably packet length, directions, radius, and logical device type, to infer 6 out of 12 encrypted NWK commands. 
However, this inference approach does not apply to APL commands.
Peekaboo~\cite{peekaboo} exploited traffic rate variations, and IoTSpy~\cite{iotspy} leveraged packet sequence features to fingerprint known IoT events of merely 3 and 5 Zigbee devices, respectively. 
In addition, Homonit~\cite{homonit} and IoTGaze~\cite{iotgaze} analyzed Zigbee event patterns to detect malicious smart home apps.
However, all these studies are confined to the identification of known events using a-priori event fingerprints. 
In contrast, ZLeaks infers event as well as device information without collecting event fingerprints for every device.
Another study, Z-IoT~\cite{ziot} employed ML to identify device type by exploiting inter-arrival-time of NWK frames and IEEE 802.15.4 Data requests of the idle device.
In contrast, ZLeaks exploits the device's periodic reporting interval and pattern (based on APL commands) to identify the device type and the \textit{device} with 99.8\% accuracy.
As evident from Table \ref{tab:table_work}, our study was conducted on the largest device set spanning 5 hubs and 19 unique Zigbee devices.

\textbf{Security of Zigbee Protocol:}
Several attacks have been demonstrated against Zigbee protocol so far, such as selective jamming~\cite{zigator}, worm chaining~\cite{iotworm}, command injection~\cite{defconlock}, replaying~\cite{replaying}, etc., with an aim to recover the Network key or make the target devices malfunction. 
Unlike ZLeaks, these attacks either rely on leaked global link key, install (QR) codes or require attacker's presence during the device's setup to identify key material.
\begin{table}[t]
\centering
\caption{ZLeaks vs. existing Zigbee based schemes for identifying Event Type (ET), Device Type (DT), Device Identity (DI) and applicability on Unknown devices (UD).}
\label{tab:table_work}
\begin{tabular}{l|c|c|l|c|c|c|c} 
\hline
\multicolumn{1}{c|}{\multirow{2}{*}{\begin{tabular}[c]{@{}c@{}}\textbf{Research }\\\textbf{Work}\end{tabular}}} & \multicolumn{2}{c|}{\textbf{Unique}}     & \multicolumn{1}{c|}{\multirow{2}{*}{\textbf{Technique (Feature)}}} & \multicolumn{4}{c}{\textbf{Identified}}                                                                                              \\ 
\cline{2-3}\cline{5-8}
\multicolumn{1}{c|}{}                                                                                           & \textbf{Hubs}      & \textbf{Devices}        & \multicolumn{1}{c|}{}                                              & \multicolumn{1}{c|}{\textbf{ET}} & \multicolumn{1}{c|}{\textbf{DT}} & \multicolumn{1}{c|}{\textbf{DI}} & \multicolumn{1}{c}{\textbf{UD}}  \\ 
\hline 
Peekaboo~\cite{peekaboo}                 & 1                  & 3                   & ML (traffic profiling)          & \tick           & \tick          &            &                 \\ 
\hline
Z-IoT~\cite{ziot}                   & 1                  & 8                   & ML (inter-arrival-times)                           &             & \tick           &            & \tick                \\ 
\hline
IoTGaze~\cite{iotgaze}                 & 1                  & 5                   & ML (event pattern)                   & \tick           &             &            &                 \\ 
\hline
IoTSpy~\cite{iotspy}                  & 1                  & 5                   & NLP (frame len (fl), direction) & \tick           &             &            &                 \\ 
\hline
Homonit~\cite{homonit}                 & 1                  & 7                   & Levenshtein Distance (fl, direction)      & \tick           &             &            &                \\ 
\hline
\multirow{2}{*}{ZLeaks} & \multirow{2}{*}{5} & \multirow{2}{*}{19} & Command Inference (metadata)                       & \tick           & \tick           & \tick          & \tick                \\ 
\cline{4-8}
                        &                    &                     & Correlation (periodic reporting)                   &            & \tick           & \tick          &                 \\
\hline
\end{tabular}
\end{table}
\section{Conclusion}
This work highlighted that the power optimization-oriented design of Zigbee protocol has destroyed the legal concept of privacy in smart homes. 
We presented ZLeaks~\cite{zleaks}, a privacy analysis tool that employs two inference techniques to demonstrate how easily a passive eavesdropper can determine in-home devices and events from the encrypted traffic, using a low-cost wireless Zigbee sniffer (TI CC2531).
The evaluation conducted on an exhaustive set of 19 unique Zigbee devices and 5 smart hubs indicates that the ZLeaks command inference technique identified unknown events and devices with 83.6\% accuracy, without using event signatures.
In addition, ZLeaks periodic reporting technique identified known devices in the absence of any user activity with 99.8\% accuracy.
Finally, we evaluated our command inference rules on a third-party capture file and identified functionality-specific APL commands with 91.2\% accuracy, irrespective of the secret keys. 
We conclude that the proposed inference attacks are impossible to prevent without making significant design changes in the Zigbee protocol. 


\begin{thebibliography}{}
\providecommand{\url}[1]{\texttt{#1}}
\providecommand{\urlprefix}{URL }
\providecommand{\doi}[1]{https://doi.org/#1}

\end{thebibliography}


\begin{thebibliography}{10}
\bibliographystyle{splncs04}
\providecommand{\url}[1]{\texttt{#1}}
\providecommand{\urlprefix}{URL }
\providecommand{\doi}[1]{https://doi.org/#1}

\bibitem{audi}
Marchal, S., Miettinen, M., Nguyen, T., Sadeghi, A.R., Asokan, N.: {AuDI: Towards autonomous IoT device-type identification using periodic  communications}. IEEE Journal on Selected Areas in Communications. (2019) %

\bibitem{profiliot}
Meidan, Y., Bohadana, M., Shabtai, A., Guarnizo, J.D., Ochoa, M., Tippenhauer, N.O., Elovici, Y.: Profiliot: A machine learning approach for iot device identification based on network traffic analysis. In: Proceedings of the symposium on applied computing. pp. 506--509. ACM, Morocco (2017)%

\bibitem{devtype1}
Miettinen, M., Marchal, S., Hafeez, I., Asokan, N., Sadeghi, A.R., Tarkoma, S.: Iot sentinel: Automated device-type identification for security enforcement in iot. In: 37th International Conference on Distributed Computing Systems. pp. 2177--2184. IEEE, USA (2017)%
  
\bibitem{event1}
Pierre Marie~Junges, J.F., Festor, O.: Passive inference of useractions through iot gateway encrypted traffic analysis. IEEE Symposium on Integrated Network and Service Management. IEEE, USA (2019)

\bibitem{pingpong}
Trimananda, R., Varmarken, J., Markopoulou, A., Demsky, B.: Packet-level signatures for smart home devices. In: Network and Distributed System Security Symposium, \textbf{10}(13), 54. NDSS, USA (2020)%

\bibitem{anybodyhome}
Copos, B., Levitt, K., Bishop, M., Rowe, J.: Is anybody home? inferring activity from smart home network traffic. In: IEEE Security and Privacy Workshops (SPW). pp. 245--251. IEEE, USA (2016)%

\bibitem{peekaboo}
Acar, A., Fereidooni, H., Abera, T., Sikder, A., Miettinen, M., Aksu, H., Conti, M., Sadeghi, A.R., Uluagac, A.S.: Peek-a-boo: I see your smart home activities, even encrypted! In: WiSec 2020 - 13th ACM Conference on Security and Privacy in Wireless and Mobile Networks. ACM, Austria (2020)%

\bibitem{homonit}
Zhang, W., Meng, Y., Liu, Y., Zhang, X., Zhang, Y., Zhu, H.: Homonit: Monitoring smart home apps from encrypted traffic. In: Proceedings of the SIGSAC Conference on Computer and Communications Security. pp. 1074--1088. ACM, Canada (2018)%
  
\bibitem{zigator}
Akestoridis, D.G., Harishankar, M., Weber, M., Tague, P.: Zigator: Analyzing the security of zigbee-enabled smart homes. In: WiSec 2020 - 13th ACM Conference on Security and Privacy in Wireless and Mobile Networks. ACM, Austria (2020). %

\bibitem{standard}
Zigbee Alliance: ZigBee Specification, 05-3474-21 edn. (2015)

\bibitem{z560}
Zigbee Alliance: 2020 and Beyond, \url{https://zigbeealliance.org/news_and_articles/zigbee-momentum/}. Last accessed Jun 2021

\bibitem{zcl}
Zigbee Alliance: Zigbee Cluster Library Specification, 07-5123-06 edn. (2016)

\bibitem{batteryreq}
Smart Home Enthusiast's Guide to ZigBee (2019), \url{https://linkdhome.com/articles/what-is-zigbee-guide}. Last accessed Jun 2021

\bibitem{zleaks}
ZLeaks, \url{https://github.com/narmeenshafqat1/ZLeaks}

\bibitem{moniotr}
Mon(IoT)r Lab, \url{https://moniotrlab.ccis.neu.edu/}. Last accessed Jun 2021

\bibitem{wshark1}
Wireshark bug, \url{https://bugs.wireshark.org/bugzilla/show_bug.cgi?id=9423}

\bibitem{crawdad}
{Zigator CRAWDAD} dataset CMU, (v. 2020-05-26), \url{https://crawdad.org/cmu/zigbee-smarthome/20200526}. Last accessed May 2021

\bibitem{tisniffer}
TI CC2531 zigbee, https://www.ti.com/product/CC2531. Last accessed Jun 2021

\bibitem{osram}
Zigbee Compliance Document of Lightify bulb (2014), \url{https://zigbeealliance.org/zigbee_products/lightify-classic-a60-rgbw/}. Last accessed Jun 2021

\bibitem{sengled}
Zigbee Compliance Document of Sengled Bulb (2018),
  \url{https://zigbeealliance.org/zigbee_products/sengled-element-3/}. Last accessed Jul 2021

\bibitem{tshark}
Tshark captures. \url{https://tshark.dev/search/pcaptable/}. Accessed May 2021

\bibitem{pyshark}
Pyshark, \url{https://github.com/KimiNewt/pyshark}. Last accessed Jun 2021

\bibitem{cve}
{US-CERT}: {CVE}. \url{http://cve.mitre.org/}. Last accessed May 2021

\bibitem{iotworm}
Ronen, E., Shamir, A., Weingarten, A.O., OFlynn, C.: Iot goes nuclear: Creating a zigbee chain reaction. In: IEEE Symposium on Security and Privacy, USA (2017)
  
\bibitem{iotnet}
Herwig, S., Harvey, K., Hughey, G., Roberts, R., Levin, D.: Measurement and analysis of hajime, a peer-to-peer iot botnet. In: Network and Distributed Systems Security Symposium. NDSS, USA (2019)%

\bibitem{laser}
Sugawara, T., Cyr, B., Rampazzi, S., Genkin, D., Fu, K.: Light commands: laser-based audio injection attacks on voice-controllable systems. In: 29th USENIX Security Symposium. pp. 2631--2648. USENIX, USA (2020)

\bibitem{padding}
Sun, Q., Simon, D.R., Wang, Y.M., Russell, W., Padmanabhan, V.N., Qiu, L.: Statistical identification of encrypted web browsing traffic. In: IEEE Symposium on Security and Privacy. pp. 19--30. IEEE, USA (2002) %

\bibitem{isend}
Leu, P., Puddu, I., Ranganathan, A., {\v{C}}apkun, S.: I send, therefore i leak: Information leakage in low-power wide area networks. In: Proceedings of 11th ACM Conference on Security \& Privacy in Wireless and Mobile Networks, Sweden (2018)%

\bibitem{sniffmislead}
Liu, X., Zeng, Q., Du, X., Valluru, S.L., Fu, C., Fu, X.:
  Sniffmislead: Non-intrusive privacy protection against wireless packet sniffers in smart homes. In: 24th International Symposium on Research in Attacks, Intrusions and Defenses (2021)

\bibitem{matter}
Matter, https://buildwithmatter.com/. Last accessed May 2021

\bibitem{iothound}
Anantharaman, P., Song, L., Agadakos, I., Ciocarlie, G., Copos, B., Lindqvist, U., Locasto, M.E.: Iothound: environment-agnostic device identification and monitoring. In: 10th International Conference on Internet of Things. ACM (2020) %

\bibitem{thangavelu19deft}
Thangavelu, V., Divakaran, D.M., Sairam, R., Bhunia, S.S., Gurusamy, M.: {DEFT: A Distributed IoT Fingerprinting Technique}. IEEE Internet of Things Journal \textbf{6}(1),  940--952 (2019)

\bibitem{cho2016fingerprinting}
Cho, K.T., Shin, K.G.: {Fingerprinting Electronic Control Units for Vehicle Intrusion Detection}. In: USENIX Security Symposium. pp. 911--927. USA (2016)

\bibitem{maldect1}
Salman, O., Elhajj, I.H., Chehab, A., Kayssi, A.: A machine learning based framework for iot device identification and abnormal traffic detection. Transactions on Emerging Telecommunications Technologies p. e3743 (2019)

\bibitem{app1}
Earlence~Fernandes, J.J., Prakash, A.: Security analysis of emerging smart home applications. In: 37th IEEE Symposium on Security and Privacy, USA (2016)

\bibitem{iotfinder}
Perdisci, R., Papastergiou, T., Alrawi, O., Antonakakis, M.: Iotfinder: Efficient large-scale identification of iot devices via passive dns traffic analysis. In: European Symposium on Security and Privacy (EuroS\&P). pp. 474--489. IEEE, virtual (2020)
  
\bibitem{ziot}
Babun, L., Aksu, H., Ryan, L., Akkaya, K., Bentley, E.S., Uluagac, A.S.: Z-iot: Passive device-class fingerprinting of zigbee and z-wave iot devices. In: IEEE International Conference on Communications (ICC). pp.~1--7. IEEE, Ireland (2020)

\bibitem{iotgaze}
Gu, T., Fang, Z., Abhishek, A., Fu, H., Hu, P., Mohapatra, P.: Iotgaze: Iot security enforcement via wireless context analysis. In: IEEE Conference on Computer Communications (INFOCOM). pp. 884--893. IEEE, virtual (2020)

\bibitem{iotspy}
Gu, T., Fang, Z., Abhishek, A., Mohapatra, P.: Iotspy: Uncovering human privacy leakage in iot networks via mining wireless context. In: IEEE 31st Annual International Symposium on Personal, Indoor and Mobile Radio Communications. pp.~1--7. IEEE, virtual (2020)%

\bibitem{defconlock}
Brown, F., Gleason, M.: Zigbee hacking: Smarter home invasion with zigdiggity. In: Black Hat USA (2019)

\bibitem{replaying}
Olawumi, O., Haataja, K., Asikainen, M., Vidgren, N., Toivanen, P.: Three practical attacks against zigbee security: Attack scenario definitions, practical experiments, countermeasures, and lessons learned. In: 14th International Conference on Hybrid Intelligent Systems. pp. 199--206. IEEE, Kuwait (2014)

\end{thebibliography}
\end{document}